\documentclass[showpacs,amsmath,superscriptaddress,preprint,aps]{revtex4-2}

\usepackage{graphicx}
\usepackage{hyperref}
\usepackage{amsmath} 
\usepackage{dcolumn}
\usepackage{bm}

\usepackage[utf8]{inputenc}
\usepackage[T1]{fontenc}
\usepackage{mathptmx}

\begin{document}

\title
{Chiral modes near exceptional points in symmetry broken H1 photonic crystal cavities}

\affiliation{Nanoscale Quantum Photonics Laboratory, RIKEN Cluster for Pioneering Research, Saitama 351-0198, Japan}
\affiliation{Quantum Optoelectronics Research Team, RIKEN Center for Advanced Photonics, Saitama 351-0198, Japan}
\affiliation{Department of Applied Physics and Physico-Informatics, Keio University, Kanagawa 223-8522, Japan}
\affiliation{Institute for Nano Quantum Information Electronics, The University of Tokyo, Tokyo 153-8505, Japan}
\affiliation{Institute of Industrial Science, The University of Tokyo, Tokyo 153-8505, Japan}
\affiliation{Research Center for Advanced Science and Technology, The University of Tokyo, Tokyo 153-0041, Japan}

\author{C.~F.~Fong}
\email[Corresponding author: ]{cheefai.fong@riken.jp}
\affiliation{Nanoscale Quantum Photonics Laboratory, RIKEN Cluster for Pioneering Research, Saitama 351-0198, Japan}
\affiliation{Quantum Optoelectronics Research Team, RIKEN Center for Advanced Photonics, Saitama 351-0198, Japan}
\affiliation{Institute for Nano Quantum Information Electronics, The University of Tokyo, Tokyo 153-8505, Japan}

\author{Y.~Ota}
\affiliation{Department of Applied Physics and Physico-Informatics, Keio University, Kanagawa 223-8522, Japan}
\affiliation{Institute for Nano Quantum Information Electronics, The University of Tokyo, Tokyo 153-8505, Japan}

\author{Y.~Arakawa}
\affiliation{Institute for Nano Quantum Information Electronics, The University of Tokyo, Tokyo 153-8505, Japan}

\author{S.~Iwamoto}
\affiliation{Institute for Nano Quantum Information Electronics, The University of Tokyo, Tokyo 153-8505, Japan}
\affiliation{Institute of Industrial Science, The University of Tokyo, Tokyo 153-8505, Japan}
\affiliation{Research Center for Advanced Science and Technology, The University of Tokyo, Tokyo 153-0041, Japan}

\author{Y.~K.~Kato}
\email[Corresponding author: ]{yuichiro.kato@riken.jp}
\affiliation{Nanoscale Quantum Photonics Laboratory, RIKEN Cluster for Pioneering Research, Saitama 351-0198, Japan}
\affiliation{Quantum Optoelectronics Research Team, RIKEN Center for Advanced Photonics, Saitama 351-0198, Japan}

\begin{abstract}
The H1 photonic crystal cavity supports two degenerate dipole modes of orthogonal linear polarization which could give rise to circularly polarized fields when driven with a $\pi$/$2$ phase difference. However, fabrication errors tend to break the symmetry of the cavity which lifts the degeneracy of the modes, rendering the cavity unsuitable for supporting circular polarization. We demonstrate numerically, a scheme that induces chirality in the cavity modes, thereby achieving a cavity that supports intrinsic circular polarization. By selectively modifying two air holes around the cavity, the dipole modes could interact via asymmetric coherent backscattering which is a non-Hermitian process. With suitable air hole parameters, the cavity modes approach the exceptional point, coalescing in frequencies and linewidths as well as giving rise to significant circular polarization close to unity. The handedness of the chirality can be selected depending on the choice of the modified air holes. Our results highlight the prospect of using the H1 photonic crystal cavity for chiral-light matter coupling in applications such as valleytronics, spin-photon interfaces and the generation of single photons with well-defined spins. 
\end{abstract}

\maketitle

\section{Introduction}\label{sec:INTRO}

The dynamics of physical systems with open boundaries that could exchange energy with their surrounding environment can be described by non-Hermitian Hamiltonians. Such systems, in general, do not have an orthogonal set of eigenstates and there exist non-trivial degeneracies known as exceptional points (EPs) at which both the eigenfrequencies and eigenstates coalesce to become one and the same. There is a surge in interest in non-Hermitian photonics and optical systems~\cite{el-ganainy2019, miri2019, parto2020} due to the relative ease to implement the complex potential required for non-Hermiticity in terms of the refractive index by incorporating gain and/or loss. For example, non-hermiticity have been successfully introduced in the whispering gallery resonators~\cite{hodaei2014,feng2014,peng2016,chen2017,chang2014,hodaei2017,miao2016} via various means such as the manipulation of excitation geometry~\cite{hodaei2014,yang2020}, heating~\cite{hodaei2017}, patterned metal depositions~\cite{feng2014,miao2016}, patterned defect scatterers~\cite{yang2021} and the usage of nanotip scatterers~\cite{peng2016,chen2017}. Intriguing phenomena have been reported in these whispering gallery resonators including single mode lasing~\cite{hodaei2014,feng2014}, enhanced sensitivity to perturbations~\cite{chen2017,hodaei2017,liu2016,wiersig2016}, directional coupling~\cite{peng2016,kim2014}, enhanced spontaneous emission~\cite{yang2021,zhong2021} and the generation of a vortex laser beam~\cite{miao2016}. 

Analogous to the whispering gallery resonator is the H1 photonic crystal (PhC) cavity due to the $C_{6v}$ rotational symmetry of the cavity. The H1 PhC cavity modes can be approximated using the cylindrical harmonics as in the whispering gallery resonator~\cite{Wiersig2011,fong2018}. A notable feature of the H1 PhC cavity is that it could support two degenerate and orthogonal linearly polarized dipole modes. When the two modes are driven with a $\pi$/$2$ phase difference, they give rise to circularly polarized cavity fields. Such a nanocavity --- given its small mode volume and prospects for high $Q$-factor --- would be an important component for chiral quantum optics~\cite{lodahl2017}, photonic circuits~\cite{coles2014, sollner2015a}, spin-nanolasers, optical sensors~\cite{troia2013} and other applications. However, these functionalities are usually hindered by fabrication errors of the nanocavity which tend to lift the mode degeneracies, making it incapable of supporting circular polarization. Previous attempts to restore mode degeneracies relied on implementing perturbation by straining the cavity~\cite{Luxmoore2012}, by purposefully designing a cavity in a “stretched lattice”~\cite{Luxmoore2011} or nanooxidation~\cite{hennessy2006}. Nonetheless, an intrinsically circularly polarized nanocavity has yet to be realized. 

In this work, based on the non-Hermitian effects in an H1 PhC cavity, we propose a scheme to achieve chiral cavity modes that could support intrinsic circular polarization. In our scheme, two air holes around the cavity are selectively modified. The modifications of the first air hole give rise to a mode splitting, while the second air hole is then modified to bring the system towards or away from the EP. From our numerical finite-difference time-domain (FDTD) simulation results, we observe the characteristic branching in the surfaces of the complex eigenfrequencies in the parameter space, indicating the presence of an EP. Near the EP, there are chiral modes with corresponding high degrees of circular polarization. The chiral modes in turn emit circularly polarized light into the farfield, shown schematically in Fig.~\ref{Fig1}(a). By selecting different pairs of air holes, modifications can be done in a controllable manner to obtain either right or left circularly polarized chiral modes. The coalescence of the modes near the EP also promises the enhancement of spontaneous emission~\cite{pick2017,takata2021} of the chiral modes.

\section{H1 photonic crystal cavity}\label{sec:H1PHC}

\begin{figure}[t]
\includegraphics[width=1.0 \columnwidth]{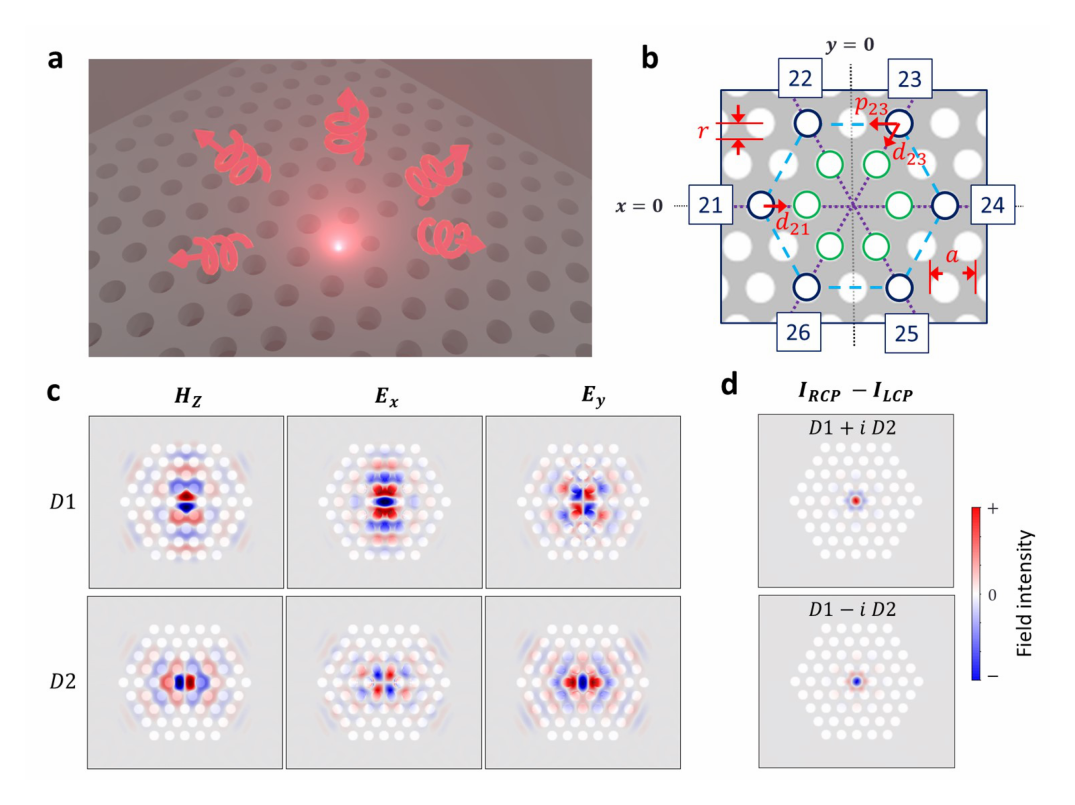}
\caption{
\label{Fig1} (a) Schematic showing the emission of circularly polarized light from a symmetry-broken H1 photonic crystal cavity. (b) Schematic of the cavity in the $xy$-plane with the key air holes and parameters labelled. See main text for the descriptions of the parameters. The first nearest air holes are outlined with green circles while the relevant second nearest air holes are outlined with black circles. The coordinate origin is at the center of the cavity. The purple dotted lines intersecting at the cavity center mark the direction of high symmetry along the $\Gamma -K$ directions while the blue dashed lines mark the hexagonal path, along which the air holes are shifted. (c) The relevant field distributions of the two dipole modes in an unmodified cavity based on 2D FDTD simulations. (d) The distribution of the difference in the field intensities with orthogonal circular polarization (spin angular momentum density, $\bm{s_d}$  $\hat{\bm{z}}$) when the dipole modes are superposed with $\pm \pi$/2 phase difference, giving predominantly RCP (LCP) field respectively. The colorbar is shared for (c) and (d).}

\end{figure}

We consider a PhC which consists of a triangular lattice of air holes with a lattice period of $a$ and radius $r$, in a slab of material of refractive index $n$. The H1 cavity is formed by removing an air hole within the lattice [Fig.~\ref{Fig1}(b)]. We first describe the relevant features and quantities of the H1 PhC cavity based on 2D FDTD simulations. Figure~\ref{Fig1}(c) shows the near-field profiles of the relevant transverse electric field components of both dipole modes D1 and D2 in which the $E_x$ and $E_y$ fields are dominant, respectively. 

Each dipole mode in the H1 PhC cavity can be thought to be constituted of two travelling wave components rotating in the opposite directions~\cite{fong2018}, analogous to whispering gallery modes~\cite{Wiersig2011}. In each dipole mode, the counter rotation of the two travelling components cancels out to result in a stationary mode. On the contrary, a superposition of the dipole modes with a $\pm\pi$/$2$ phase difference essentially recovers one of the travelling modes that rotates in the clockwise (counterclockwise) direction, producing a chiral mode with a dominant right (left) circularly polarized field [Fig.~\ref{Fig1}(d)]. This phenomenon is known as spin-momentum locking, in which the direction of rotation of the travelling mode is directly correlated to the handedness of its chirality.

In order to quantify chirality, one could first calculate the spin angular momentum density~\cite{Aiello2015},  $\bm{s_d} =$ Im $(\epsilon_o \epsilon_r \bm{E}^* \times \bm{E} + \mu_o \mu_r  \bm{H}^* \times \bm{H})/4$ which can be regarded as the difference in the intensity of right (RCP) and left circularly polarized (LCP) fields. $\bm{E}$ ($\bm{H}$) represents the electric (magnetic) field vectors, the asterisk (*) indicates complex conjugation and Im() means taking the imaginary part. $\epsilon_o$ ($\epsilon_r$) and $\mu_o$ ($\mu_r$) are the vacuum (relative) permittivity and vacuum (relative) permeability respectively. The degree of circular polarization (DCP) can then be obtained by dividing $\bm{s_d}$ with the total field energy density, $W=(\epsilon_o \epsilon_r \bm{E}^* \cdot \bm{E} + \mu_o \mu_r  \bm{H}^* \cdot \bm{H})/4$, giving a value between $\pm1$. Since the cavity modes are transverse electric in nature with the electric fields oscillating in the plane of the PhC slab, $\bm{s_d}$ points in the out-of-plane $\pm \hat{\bm{z}}$ direction and thus we only consider the $z$-component of $\bm{s_d}$ and its associated DCP. Furthermore, as the dominant electric field components and the $z$-component of $\bm{s_d}$ have the strongest intensity at the cavity center, the DCP at the cavity center can be taken as a representative measure of the chirality of the cavity.

2D FDTD simulations are performed using a PhC cavity with  an overall hexagonal shape so that it is consistent with the $C_{6v}$ symmetry of the cavity in order to avoid or reduce any unintended perturbation to the cavity modes. It is important to note that the PhC is of a finite size with lateral loss such that it forms a non-Hermitian open boundary system. Since the two modes need to be overlapping to support chirality, we ensure that the spectral widths of the modes are sufficiently broad by using a  $9 \times 9$ air hole lattice. The FDTD simulations are performed using the opensource package MEEP~\cite{oskooi2010}, with parameter values $a = 300$ nm, $r/a = 0.3$, and a grid resolution of ${\sim} a$/16 (54 pixels/micron). We consider a material of $n$ = 3.4, for example GaAs. A Gaussian-pulse point current source with a sufficient spectral width is placed at the center of the cavity as the excitation source. Given these parameters, the dipole modes are obtained at a normalized frequency of $a/\lambda_{cav}\sim 0.244$. The $Q$-factor, which is the ratio of the mode center frequency to its linewidth, is ${\sim} 450$. For an unmodified cavity, the modes are not completely degenerate, with a splitting of ${\sim} 1.5$ linewidth for the parameters given above. The non-degeneracy is possibly due to the discretization of the simulation grid and it persists even at higher grid resolutions. Nonetheless, this non-degeneracy poses no significant issues to our scheme. On the contrary, it presents us with a system that more closely resembles an actual sample which tend to have non-degenerate modes due to fabrication errors. 

\section{Surfaces of complex eigenfrequencies and exceptional point}\label{sec:EP}

\begin{figure*}[t]
\centering
\includegraphics[width=0.95 \textwidth]{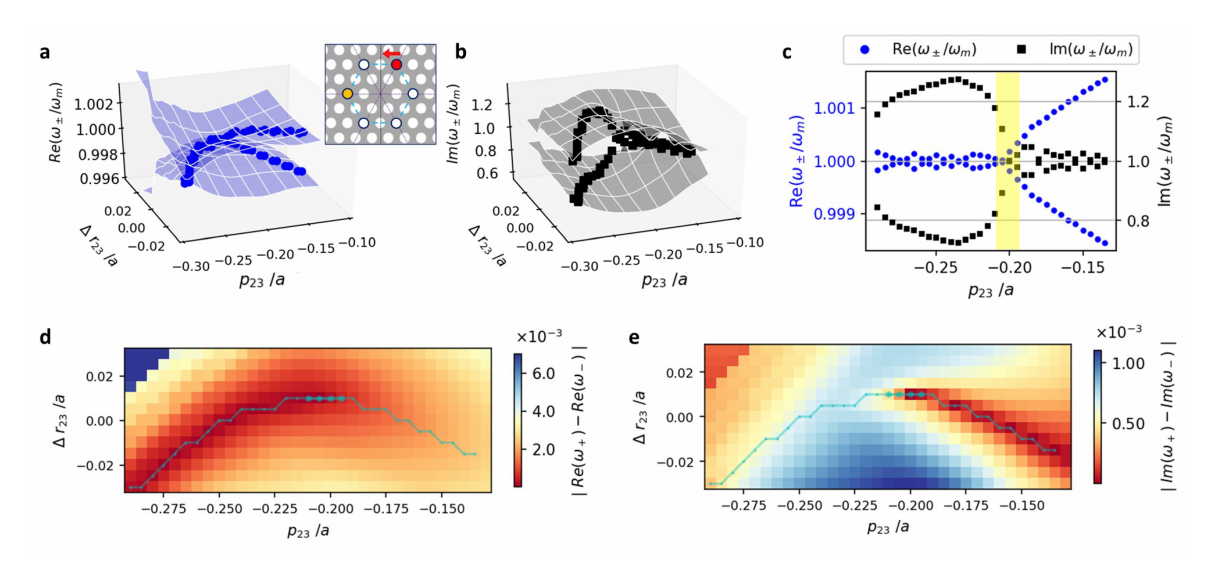}
\caption{
\label{Fig2} Surfaces of the (a) real and (b) imaginary eigenfrequencies normalized to their midpoints $\omega_m=(\omega_++\omega_- )/2$. The points of interest that gives the characteristic branching associated with the presence of an EP is marked on the surfaces in blue dots and black squares for the real and imaginary eigenfrequencies respectively.  (Inset) Schematic indicating the air hole with fixed modifications only (orange circle) and the air hole with both fixed and parameter sweep modifications (red circle). The red arrow indicates the direction that the air-hole is being shifted along the hexagonal path. (c) The branching of the real (blue dots) and imaginary (black squares) frequencies plotted against $p_{23}$. The region where the EP is expected to occur is highlighted in yellow. Color plots showing the absolute difference in the (d) real and (e) imaginary eigenfrequencies within the parameter space spanned by $\Delta r_{23}$ and $p_{23}$. The cyan line traces along the branching of the eigenfrequencies. The region where the EP is expected to occur indicated with bigger data markers. 
}
\end{figure*}

In our scheme to achieve a chiral cavity, two second nearest air holes to the cavity in the $\Gamma-K$ directions are modified to give the necessary perturbation to the modes. In principle, our scheme could work by modifying either the first or the second nearest air holes to the cavity. The first nearest  air holes are in the direct vicinity of the modes and thus even small modifications to these air holes could result in excessive perturbations. Despite being further away from the cavity, modifications to the second nearest air holes could still provide sufficient perturbation. In addition, certain second nearest air holes are positioned closer to the antinodes of either one of the two dipole modes. As such, each modified air hole would predominantly perturb one of the dipole modes. For these reasons, our scheme is based on the second nearest air holes as they allow for better control and intuitive understanding over the effects of their modifications. 

The relevant second nearest air holes for our scheme are labelled as 21 to 26 in Fig.~\ref{Fig1}(b). The radii and the positions of two air holes are modified at a time. The positions of the air holes will be shifted along the direction of high symmetry [purple lines in Fig.~\ref{Fig1}(b)] towards (negative shift) or away (positive shift) from the cavity. The air holes could also be shifted along the hexagonal path around the cavity [blue dashed lines in Fig.~\ref{Fig1}(b)] with clockwise (counterclockwise) shift being the positive (negative) direction. The hexagonal path is defined by the distance from the cavity center to the center of the relevant air hole after the air hole has been shifted along the direction of high symmetry. The change in the air hole radius, shift along the direction of high symmetry and shift along the hexagonal path will be labelled as $\Delta r$, $d$ and $p$, respectively, followed by the hole number in subscript e.g., $\Delta r_{21}$. The different cases described in this work will be referred to in accordance with the pair of air holes that are being modified, for example h21h23 refers to the case in which air holes 21 and 23 are modified.

We begin with the h21h23 case, where the following fixed modifications are applied to the air holes: radius of air hole 21 is enlarged ($\Delta r_{21} = +0.03a$) and shifted towards the cavity ($d_{21}=-0.1a$), while air hole 23 is also shifted towards the cavity ($d_{23}=-0.20a$). $p_{23}$ and $\Delta r_{23}$ are then varied over a range of values for the parameter sweep simulations. In particular, air hole 23 is shifted towards $y = 0$ i.e., negative $p_{23}$ values. FDTD simulation is performed for each combination of $p_{23}$ and $\Delta r_{23}$ to extract the complex eigenfrequencies of the two modes $\omega_{\pm}$, as well as the temporal evolution of $\bm{s_d}$ $\hat{\bm{z}}$, $W$, and the DCP. 

Presented in Fig.~\ref{Fig2}(a) and~\ref{Fig2}(b) are the surfaces of the extracted real and imaginary eigenfrequencies, Re($\omega_{\pm}$) and Im($\omega_{\pm}$), respectively, normalized to their midpoints $\omega_m=(\omega_++\omega_- )/2$ within the parameter space spanned by $\Delta r_{23}$ and $p_{23}$. The surfaces of the eigenfrequencies exhibit characteristics of a non-Hermitian system. For both Re($\omega_{\pm}$) and Im($\omega_{\pm}$), there are particular regions where the surfaces come close together, indicating that the eigenfrequencies are close to degeneracy. The points of interest are marked with blue dots (black squares) for the real (imaginary) eigenfrequencies and their values plotted in Fig.~\ref{Fig2}(c). The resulting plot shows the branching in the eigenfrequencies with $p_{23}$ as the “tuning parameter”. To show where these points of interests lie within the parameter space, Fig.~\ref{Fig2}(d) and~\ref{Fig2}(e) present a different visualization of the eigenvalue surfaces, namely, the distributions of the absolute difference in the real and imaginary eigenfrequencies, respectively. In the range of $p_{23}=-0.30a$ to $-0.20a$, Re($\omega_{\pm}$) are close to being degenerate over a range of $\Delta r_{21}$ values, while  Im($\omega_{\pm}$) are split into two branches. At around $p_{23}=-0.20a$,  Re($\omega_{\pm}$) begins to branch while the two branches of Im($\omega_{\pm}$) merge.

The branching of eigenfrequencies indicate the presence of an EP within the parameter space defined by all the five parameters that we consider here --- $d_{21}$, $\Delta r_{21}$, $d_{23}$, $\Delta r_{23}$ and $p_{23}$  --- and possibly other parameters. At the exact EP,  the branch points of Re($\omega_{\pm}$) and Im($\omega_{\pm}$) will coincide. In our simulation reults, the branch points of Re($\omega_{\pm}$) and Im($\omega_{\pm}$) do not coincide exactly which could be related to the non-degeneracy of the dipole modes of the unmodified H1 PhC cavity. Although the location of the EP becomes less well-defined, the EP is expected to occur in the vicinity of $p_{23}=-0.20a$ as marked by the yellow region in Fig.~\ref{Fig2}(c).

\section{Chiral modes}\label{sec:chiralmodes}

To determine the chirality of the modes, we perform FDTD simulations with long simulation time of about 2000 wave period propagation for each combination of parameter values. The simulation time is dependent on the cavity field lifetime which is associated with the $Q$-factor. The simulation time --- of more than five times the cavity field lifetime --- used here is to ensure that the cavity field evolves sufficiently to reflect its intrinsic polarization. The DCP at the cavity center at every fixed timestep is recorded during each simulation. From the temporal evolution of the DCP at the cavity center, we calculate its time-averaged value, $\langle$DCP$\rangle$, taking care to consider only the temporal evolution of DCP after the excitation source has been turned off. 

\begin{figure*}[t]
\includegraphics[width=0.9 \textwidth]{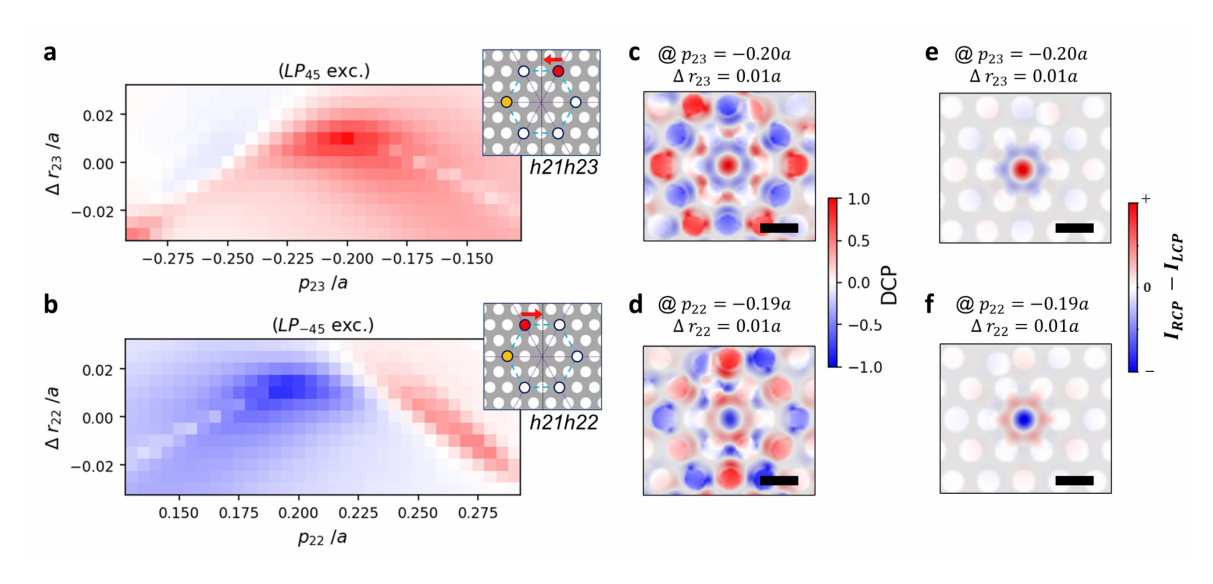}
\caption{
\label{Fig3}Distribution of $\langle$DCP$\rangle$ in the parameter space for (a) the h21h23 case under LP$_{45}$ excitation and (b) the h21h22 case under LP$_{-45}$ excitation. (Insets) Schematics indicating the modified air holes. (c, d) The distribution of the DCP around the cavity at ${\sim}1700$ wave period propagation with the parameter values where the $\langle$DCP$\rangle$ is maximally right and left circular polarized for the two respective cases. The scale bar represents one lattice period. The color bar is shared for (a-d). (e, f) The distribution of the difference in the field intensities (spin angular momentum density) where the $\langle$DCP$\rangle$  is maximally right and left circular polarized for the RCP and LCP chiral cases, respectively. The field intensity and thus field energy density is centered in the cavity as expected. The DCP at the center of the cavity can be taken as a measure of chirality. The color bar is shared for (e) and (f).
}
\end{figure*}

Figure~\ref{Fig3}(a) show the distribution of $\langle$DCP$\rangle$ under linearly polarized excitation oriented at $45^{\circ}$ relative to the $x$-axis (LP$_{45}$) for the h21h23 case. Despite the linearly polarized excitation, we observed RCP chiral eigenmodes near the EP, achieving the largest $\langle$DCP$\rangle$ of 0.94 at parameters $p_{23}=-0.20a$ and $\Delta r_{23}=0.01a$. At the chiral mode, the cavity fields evolve temporally from its initial linearly polarized state --- as determined by the excitation polarization --- to its intrinsic polarization. A temporal slice of the spatial distribution of DCP around the cavity at ${\sim}$1700 wave period propagation at the parameters with the largest $\langle$DCP$\rangle$ is presented in Fig.~\ref{Fig3}(c). The DCP at the center of the cavity at this instant can be taken as a measure of the intrinsic polarization, indicating that the cavity modes are intrinsically RCP with a DCP of 0.97. While there are other regions around the cavity with significant DCP, the field intensity [Fig.~\ref{Fig3}(e)] at these regions is negligible and thus irrelevant to the mode chirality.

A different example with LCP chiral modes is the h21h22 case [Fig.~\ref{Fig3}(b)]. The fixed modifications to air holes 21 and 22 are similar to that in the h21h23 case: $\Delta r_{21}=+0.02a$, $d_{21}=-0.10a$ and $d_{22}=-0.20a$. For the parameter sweep, the air hole 22 is shifted towards $y = 0$ by applying positive $p_{22}$ values. The distribution of the eigenfrequencies and the branching is largely similar to that of the h21h23 case but flipped about the  vertical axis (not shown). Performing the simulations under LP$_{-45}$  excitation, the $\langle$DCP$\rangle$ in the parameter space indicates LCP chiral eigenmodes in the vicinity of the EP [Fig.~\ref{Fig3}(d)], achieving the largest $\langle$DCP$\rangle$ of $-0.80$ at parameters $p_{22}=-0.19a$ and $\Delta r_{22}=0.01a$. The corresponding temporal slice of the DCP distribution around the cavity [Fig.~\ref{Fig3}(d)] gives an intrinsic polarization of $-0.83$, with the field intensities similarly centered at the cavity [Fig.~\ref{Fig3}(f)].

Despite using modifications of the same magnitude in the air holes in both the h21h22 and h21h23 cases, the difference in the distribution of $\langle$DCP$\rangle$ in the two cases could be due to the interaction between the two modified air holes via scattering~\cite{Wiersig2011}. Such interactions between air holes are more likely to occur for the h21h22 case since the modified air holes are closer to each other. These findings suggest that the air hole modifications of the h21h23 and h21h22 cases need to be optimized separately to obtain the intended chiral modes with high DCP.

As an alternative to the h21h22 case, air holes 22 and 24 can be modified instead, i.e., a mirror reflection of the h21h23 case about the $y$-axis, which gives LCP chiral modes with the same magnitude in the DCP as that in the h21h23 case. This shows the versatility of our scheme: when a chiral mode is found, the mode with the opposite chirality can be obtained simply by symmetry considerations, namely by reflection of the PhC cavity about the $x$- or $y$-axis. 

We confirmed that the eigenmodes near the EP remain RCP or LCP for the respective cases presented above regardless of the excitation polarization, providing further support that the chirality is indeed an intrinsic property. The excitation polarization will, however, affect the initial polarization of the cavity modes and the subsequent temporal evolution of the cavity field polarization, giving rise to varying $\langle$DCP$\rangle$. The $\langle$DCP$\rangle$ values are slightly lower than the intrinsic DCP values since the cavity modes evolve from the initial linearly polarized state with zero DCP to its intrinsic chiral state over some time. We find that for linearly polarization excitation, orientation at $45^{\circ}$($-45^{\circ}$) give the largest achievable $\langle$DCP$\rangle$ for the RCP (LCP) chiral modes (see Appendix~\ref{sec:toymodel} for further details). Our choice of excitation polarization is based on this observation, which helps to simultaneously highlight the intrinsic nature of the chirality as well as to show the large DCP of the chiral modes.

As mentioned in section \ref{sec:H1PHC}, the dipole modes can be thought of as a superposition of rotating components. As the eigenmodes approach the EP, they coalesce and eventually become one and the same at the EP. In this regime, the eigenmodes are no longer stationary. The asymmetric coherent backscattering between the two eigenmodes causes unbalanced amplitudes of the constituent counter-rotating components~\cite{Wiersig2011}. As a consequence, only one of the constituent rotating components is dominant and both the eigenmodes co-rotate in the same direction. Clockwise (counterclockwise) rotating mode will result in RCP (LCP) mode due to spin-momentum locking. Such clockwise (counterclockwise) rotation of the field profiles is indeed observed in the FDTD simulations for chiral modes of the h21h23 (h21h22) cases (see Appendix~\ref{rotatingmodes}). The choice of modified air holes --- 22 or 23 --- will mainly induce either a “forward or backward” backscattering, allowing one to select a dominant rotating mode and thus the handedness of the chirality (see Appendix~\ref{sec:selectchirality} for further details).

\section{3D FDTD simulation} \label{sec:3DFDTD}

\begin{figure*}[t]
\includegraphics[width=0.88 \textwidth]{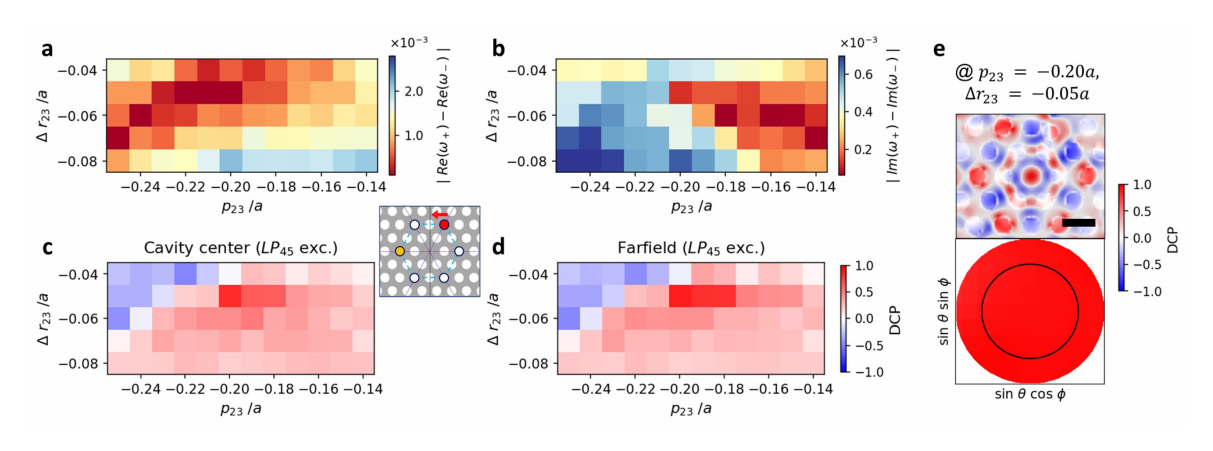}
\caption{
\label{Fig4}Simulation results of 3D h21h23 case. Distribution of the absolute difference in the (a) real and (b) imaginary eigenfrequencies within the parameter space. Color plots showing (c) $\langle$DCP$\rangle$ at cavity center and (d) the farfield $\langle$DCP$\rangle$ within NA = $0.65$ in the parameter space. The colorbar is shared for (c-d). (e) Temporal slices of the DCP around the cavity (top) and in the farfield (bottom) ${\sim}2000$ wave period propagation for parameters $p_{23}=-0.20a$ and $\Delta r_{23}=-0.05a$. At this instant, the DCP at the center of the cavity approaches 0.92 and the mean DCP within NA = 0.65 (black circle) approaches 0.94. The colorbar is shared for both plots. The scale bar indicates one lattice period.  
}
\includegraphics[width=0.88 \textwidth]{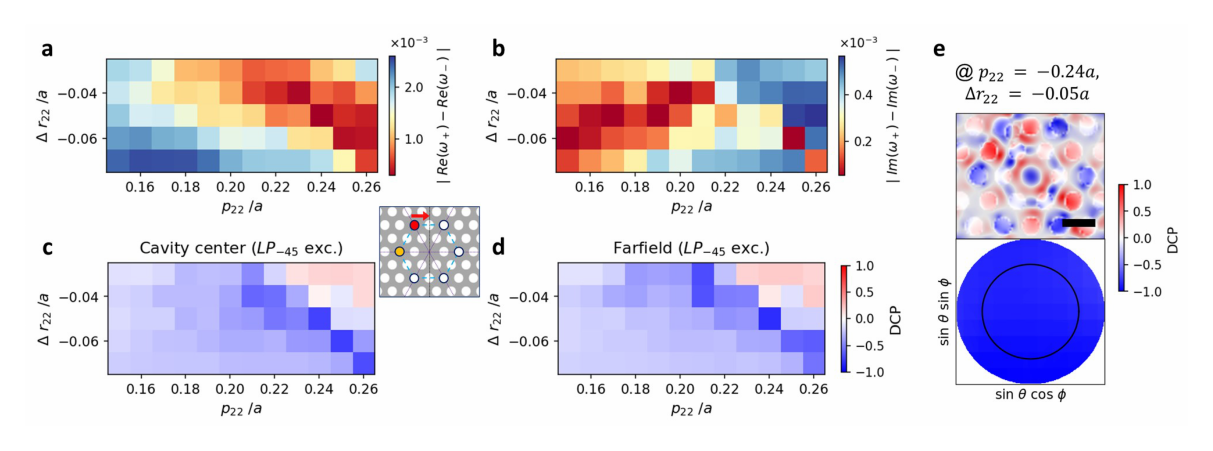}
\caption{
\label{Fig5}Simulation results of 3D h21h22 case. Distribution of the absolute difference in the (a) real and (b) imaginary eigenfrequencies within the parameter space.  Color plots showing (c) $\langle$DCP$\rangle$ at cavity center and (d) the farfield $\langle$DCP$\rangle$ within NA = $0.65$ in the parameter space. (e) Temporal slices of the DCP around the cavity (top) and in the farfield (bottom) ${\sim}2000$ wave period propagation for parameters $p_{22}=-0.24a$ and $\Delta r_{22}=-0.05a$. At this instant, the DCP at the center of the cavity approaches $-0.83$ and the mean DCP within NA = 0.65 (black circle) approaches $-0.90$. The colorbar is shared for both plots. The scale bar indicates one lattice period. 
}
\end{figure*}

To simulate practical 3D devices, we consider a PhC with the same $9 \times 9$ air hole lattice,  lattice period and air hole radius as in the 2D simulations, with an additional dimension of slab thickness set to be $0.5a$. A grid resolution of ${\sim}a/17$ ($57$ pixels/micron) is used in the simulations. The $Q$-factor of an H1 PhC cavity mode tends to be low, as modifications to the nearest air holes are usually required to achieve a high $Q$-factor~\cite{Takagi2012}. In order to show that our scheme is compatible with other modifications, we changed the radius of all the first nearest air holes ($r_1$) from $0.3a$ to $0.28a$, increasing the $Q$-factor from ${\sim}200$ to ${\sim}320$. The dipole modes of the $r_1$-modified H1 PhC cavity are at  $a/\lambda_{cav} {\sim}0.294$ with a mode volume of ${\sim}0.35(\lambda/n)^{3}$. The dipole modes are not completely degenerate with a splitting of ${\sim}0.5$ times the linewidth.

To achieve chiral modes, we apply the following modifications to the selected second nearest air holes for the h21h23 case: $\Delta r_{21}=+0.01a$, $d_{21}=-0.115a$, and $d_{23}=-0.225a$. Figure~\ref{Fig4}(a) and~\ref{Fig4}(b) shows the absolute difference of the Re($\omega_{\pm}$) and Im($\omega_{\pm}$) respectively, with the distributions reflecting the complex eigenfrequency surfaces with the characteristic branching indicating the presence of an EP as expected. 

In 3D simulations, we are able to extract the farfield emission. The farfield --- in the spherical coordinate --- is calculated from the nearfield distribution via Fourier transform~\cite{Vuckovic2002a,Kim2006a}. In addition to the DCP at the cavity center, we monitor the temporal evolution of the DCP of the farfield emission. At every fixed timestep, the mean farfield DCP within a numerical aperture (NA) $\leq$ $0.65$ is calculated and recorded. The choice of NA $\leq$ $0.65$ is to reflect realistic experimental conditions based on the commonly available and often-used microscope objective lenses.

It is found that the temporal evolution of the DCP at the cavity center and the mean farfield DCP follow each other closely, especially in terms of the trend though the values may differ slightly. The time-averaged DCP and mean farfield DCP in the parameter space obtained under LP$_{45}$ excitation is presented in Fig.~\ref{Fig4}(c) and~\ref{Fig4}(d) respectively.  RCP chiral modes are observed near the EP as expected, with the $\langle$DCP$\rangle$ reaching 0.82 at the cavity center and 0.88 in the farfield at $p_{23}=-0.20a$ and $\Delta r_{23}=-0.05a$. Figure~\ref{Fig4}(e) show the temporal slices of the DCP around the  cavity (top) and in the farfield (bottom) at ~${\sim}$2000 wave period propagation for parameters $p_{23}=-0.20a$ and $\Delta r_{23}=-0.05a$. The farfield emission is RCP throughout. At this instant, the DCP at the center of the cavity approaches 0.92 and the mean DCP within NA = 0.65 (black circle) approaches 0.94 indicating the intrinsic DCP of the chiral modes. 

For the 3D h21h22 case, slightly different fixed modifications are used: $\Delta r_{21}=-0.01a$, $d_{21}=-0.145a$ and $d_{22}=-0.23a$. We obtained LCP chiral modes with $\langle$DCP$\rangle$ of $-0.73$ and $-0.82$ at the cavity center and in the farfield, respectively, at $p_{22}=-0.24a$ and $\Delta r_{22}=-0.05a$ [Fig.~\ref{Fig5}]. The intrinsic DCP approaches $-0.83$ at the cavity center and $-0.90$ in the farfield.

To achieve chiral modes in this scheme, one only needs to find suitable parameter values to bring the system close to EP without having obtain the exact EP. Strictly speaking, only the eigenmodes near the EP exhibit significant chirality and the chirality goes to zero at the EP~\cite{peng2016}. As such, chiral modes can be achieved under less stringent conditions, further highlighting the practicality of our scheme. In Appendix~\ref{sec:toymodel}, we present a discussion of a two-mode approximation model that supports our key findings in the FDTD simulations. 

Comparing simulations results for the $r_1$-modified and unmodified H1 PhC cavity (results not shown), we find that the $r_1$-modified PhC cavity requires larger modifications to the second nearest air holes in order to achieve chiral modes. Using the h21h23 case as an example, with $r_1$ modified, holes 21 and 23 need to be shifted closer to the cavity and the radius of hole 23 needs to be reduced by larger magnitude. We also found that larger changes to the first nearest holes to increase the $Q$-factor will require larger modifications to the second nearest air holes to obtain chiral modes. Should the required modifications to the second nearest air holes become so significant such that the structure is rendered impractical --- for example, adjacent air holes are in contact or that the air holes become too small for nanofabrication --- one could instead modify two selected first nearest air holes to achieve chiral modes. As such, by applying all necessary modifications to the first nearest air holes, one could achieve a high-Q chiral cavity. 

A PhC nanocavity confines optical fields within a small mode volume facilitating strong light-matter coupling as well as the Purcell effect which describes the enhanced spontaneous emission. The local density of states near the EP is altered due to the non-orthogonality of the modes which could in turn give additional enhancement to the spontaneous emission~\cite{pick2017,pick2017a,takata2021}. This additional enhancement can be improved by increasing the $Q$-factor or by introducing material gain~\cite{pick2017} which are both applicable to the chiral PhC cavity, highlighting the prospects of using such chiral PhC cavity for device applications.

\section{Conclusion} \label{sec:conclusion}

We have presented a scheme to design intrinsically circularly polarized chiral H1 PhC cavities. In this scheme, the symmetry of the cavity is broken intentionally by modifying two of the second-nearest air holes to the cavity in the $\Gamma -K$ directions. The modifications induce an asymmetric coherent backscattering between the two eigenmodes, which is a non-Hermitian process. As a result, within the parameter space of the modified air holes, there are EPs at which the eigenfrequencies and the eigenmodes coalesce. In the vicinity of the EP, the eigenmodes are no longer stationary but are co-rotating modes with an overall direction. This in turn gives rise to chiral modes in which the DCP is correlated with the rotation direction of the mode profile. From the 3D FDTD simulation results, we show that a practical device is achievable in which both the nearfield and farfield emission exhibit near unity DCP. The handedness of the chirality can be controlled by modifying selected air hole pairs. 

The PhC nanocavity localizes the optical fields in a small mode volume while being able to maintain a relatively high $Q$-factor and thus capable of achieving a high $Q$-to-mode volume ratio which is important for quantum information technologies~\cite{Hennessy2007}. In addition, incorporating chirality in the form of circular polarization in our proposed H1 PhC nanocavity will provide additional degrees of freedom for optical control and information processing. Our proposed chiral PhC nanocavity expands the toolbox of exceptional point photonics and will complement existing chiral whispering gallery mode resonators ~\cite{peng2016, kim2014, zhong2021}, chiral photonics structures~\cite{lodahl2017, parappurath2020, mehrabad2020} as well as metamaterials~\cite{wang2016, konishi2020} for extended functionalities. By exploiting the favorable properties of the chiral H1 PhC nanocavity, one could expect the further development of chiral photonics application such as valleytronics with 2D materials~\cite{gong2018,rong2020}, spin-photon interfaces~\cite{Carter2013,Lodahl2015}, and the generation of single photons with well-defined spins~\cite{kan2020}.

\appendix

\section{Selecting chirality}\label{sec:selectchirality}

In the case of a microdisk or microring resonator, two scatterers can be placed close to the edge of the resonator to induce the necessary backscattering to bring about EPs~\cite{peng2016,yang2021,Wiersig2011}. One of the scatterers is usually fixed, for example at the field antinode of the whispering gallery mode and the position of the second scatterer is varied. The backscattering conditions to achieve an EP are met when the second scatterer is rightly positioned close to an antinode and EPs would occur periodically as the second scatterer is moved across consecutive antinodes. In addition, the system would alternate between dominant forward and backward backscattering, therefore the associated handedness of the chirality would also switch periodically. 

Consequently, it should be possible to selectively induce an EP of a specific chirality by a judicious placement of scatterers relative to the antinodes of the mode profiles. Note that the antinode here refers to the antinode of a nominally unperturbed mode profile as significant perturbation near the EP could strongly affect the mode profile and reduce the visibility of the antinodes~\cite{peng2016,Wiersig2011}.

The h21h22 and h21h23 H1 PhC cavity cases described in the main text are consistent with the previously reported observations in the whispering gallery resonators. In the H1 PhC cavities, the modified air holes behave as the scatterers. Only one EP exists within the parameter space in each of the two cases. The resulting chiralities of the eigenmodes near the EPs are of opposite handedness in the two cases depending on whether the modified air hole is to the left or the right of the antinode of the D1 dipole mode profile. 

\begin{figure*}[t]
\includegraphics[width=\textwidth]{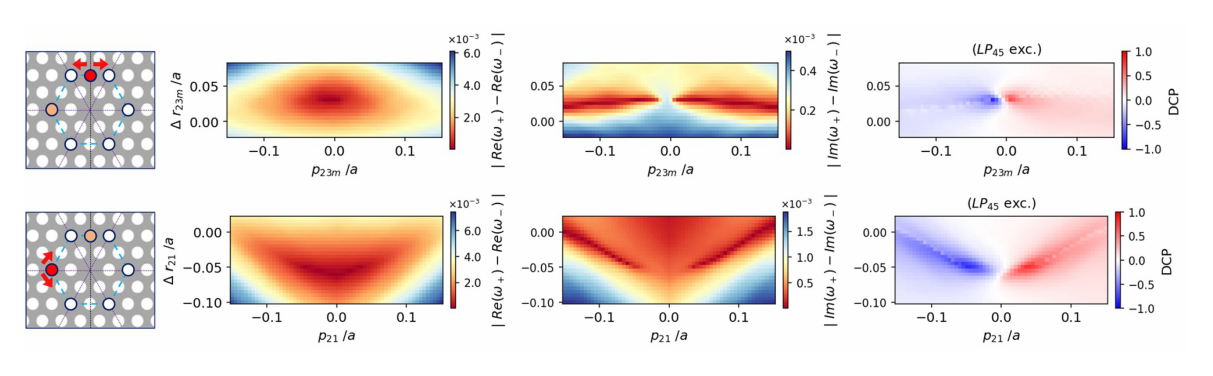}
\caption{
\label{Fig6}h21h23m cases in which parameter sweep is performed on air hole 23m (top row) and air hole 21 (bottom row) respectively. The color plots present the absolute difference in the eigenfrequencies and the ⟨DCP⟩ at cavity center as  labelled. A $45^{\circ}$ linearly polarized source is used for excitation. In both cases, the distribution of the absolute difference in the eigenfrequencies indicate the presence of two EPs. The chirality associated with each EP is of opposite handedness, giving RCP (LCP) eigenmodes when respective air holes are shifted in the clockwise (counterclockwise) direction along the hexagonal path. For the simulations in the top row, the fixed air hole modifications are $\Delta r_{21}=+0.02a$, $d_{21}=-0.08a$ and $d_{23m}=-0.17a$. For the bottom row, the fixed air hole modifications are given by $\Delta r_{23m}=+0.04a$, $d_{23m}=-0.10a$ and $d_{21}=-0.22a$.
}
\end{figure*}

To further verify this idea, we perform simulations in which air hole 21 and the air hole between 22 and 23 --- which we refer to as air hole 23m --- are modified. Both the air holes are originally aligned to the antinodes of the D1 and D2 dipole mode profiles. As such, moving either one of the air holes along the hexagonal path should, in principle, give 2 EPs with associated chiralities of opposite handedness within the parameter space. In the simulations, we use LP$_{45}$ excitation in the simulations to demonstrate the independence of the mode chirality on the excitation polarization. The expected behavior is indeed observed in the 2D FDTD simulation results [Fig.~\ref{Fig6}]. Under co-circular polarization excitation, the $\langle$DCP$\rangle$ of the chiral modes approaches $0.95$.

For the h21h22 case described in the main text [Fig.~\ref{Fig3}(b)], there is a region around $p_{22}=0.25a$ to $0.275a$ in which the modes have a modest degree of RCP instead of LCP. The appearance of such a region suggests the onset of formation of the second EP with the opposite chirality for the particular set of $\Delta r_{21}$, $d_{21}$ and $d_{22}$ parameter values used. Such a corresponding opposite chirality region is also present, though much less prominent, in the h21h23 case [Fig.~\ref{Fig3}(a), $p_{23}=-0.275a$ to $-0.25a$]. By further modifying the air hole parameter values, it is possible to clearly observe two EPs within the parameter space.

\section{Two-mode approximation model} \label{sec:toymodel}

\begin{figure*}[t]
\includegraphics[width=0.80 \textwidth]{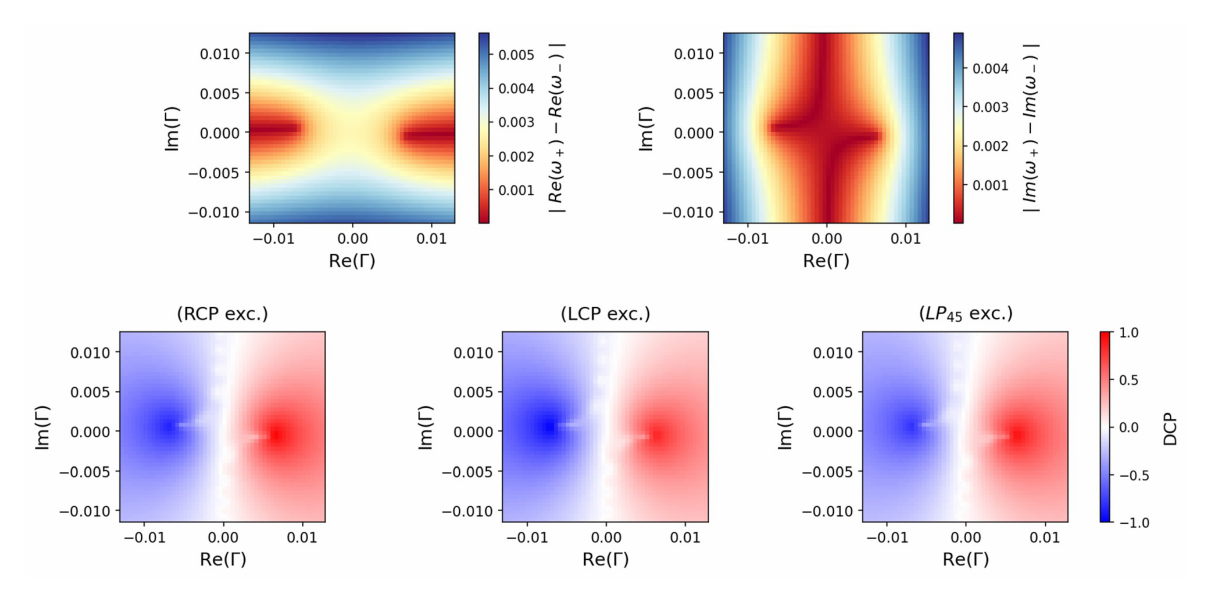}
\caption{
\label{Fig7}(Top row) The distribution of the absolute differences in the real and imaginary eigenfrequencies within the parameter space. (Bottom row) The distribution of the $\langle$DCP$\rangle$ under RCP, LCP and linearly polarized excitation oriented at $45^{\circ}$. The colorbar is shared for the plots in the bottom row. The following values are used for these modelling results: $\omega_1=1.0027-0.0022i, \omega_2=1-0.0020i, f_1=0.025, f_2=0.04, s_1=0.195i$ and $s_2=0.200i$.    
}
\end{figure*}

\begin{figure*}[t]
\includegraphics[width=0.75 \textwidth]{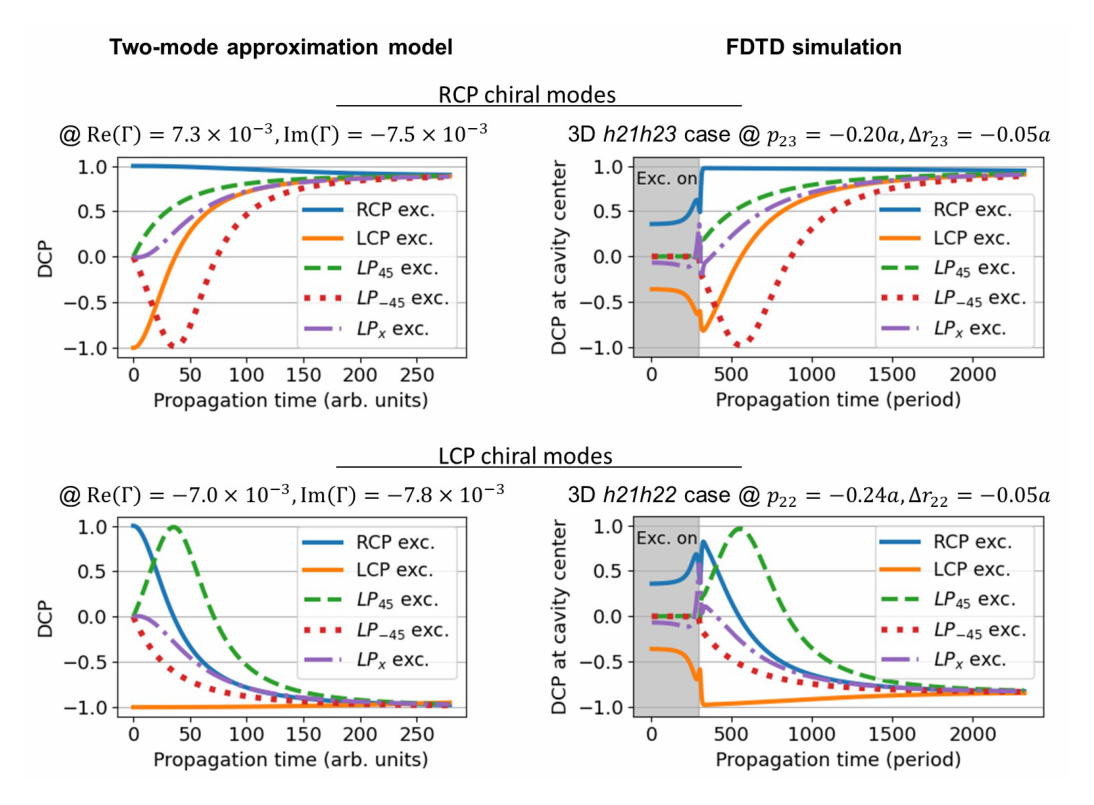}
\caption{
\label{Fig8}Temporal evolution of DCP calculated using two-mode approximation model (left column) and FDTD simulations (right column). The top (bottom) row compares the temporal evolution of DCP under different excitation polarization for the corresponding RCP (LCP) chiral modes obtained in the two-mode approximation and 3D FDTD simulations. The relevant parameter values are noted on top of each plot. The grey areas in the plots of the FDTD simulation results indicate the period of time when the excitation is on. The temporal evolution of DCP under vertical linear polarization, LP$_y$ excitation (not shown) is exactly the same as that under horizontal linear polarization, LP$_x$ for both the two-mode approximation model and FDTD simulations. The temporal evolution of DCP in the farfield from the FDTD simulations is largely similar to that at the cavity center and thus not presented here.     
}
\end{figure*}

For a non-Hermitian PhC nanocavity system that is sufficiently close to the EP, it can be described by a 2-dimensional Hamiltonian associated with the two coalescing states~\cite{heiss2012}. To approximate a H1 PhC cavity, we can define a Hamiltonian as follows, in which the basis states are the $E_x$ and $E_y$ electric fields of the two respective linearly polarized dipole modes: 

\[
H(\Gamma) =
\begin{bmatrix}
\omega_1 & 0 \\ 0 & \omega_2
\end{bmatrix}
+\Gamma
\begin{bmatrix}
f_1 & s_1 \\ s_2 & f_2
\end{bmatrix}
\]

The second term on the right-hand side can be thought of as the perturbation. The terms $\omega_{1,2}$ and $f_{1,2}$ determine the non-interacting resonance frequencies $\omega_{1,2}+\Gamma f_{1,2}$. The terms $s_1 (s_2)$ representing the coherent backscattering from mode D1 to D2 (D2 to D1) and $\Gamma$ being the tuning parameter. For $s_{1,2}\neq 0$, the eigenfrequencies are

\begin{equation} \label{eqn:eigenfreq}
\begin{split}
\omega_\pm(\Gamma) = & 1/2 \Bigl( \omega_1+\omega_2 +\Gamma(f_1+f_2 ) \\  
& \pm \sqrt{(f_1-f_2 )^2+4 s_1 s_2 (\Gamma-\Gamma_1 )(\Gamma-\Gamma_2 )} \Bigr)             
\end{split}
\end{equation}

where $\Gamma_1$ and $\Gamma_2$ are the values at which the EP occurs, given by $\Gamma_1=\frac{-i(\omega_1-\omega_2 )}{i(f_1-f_2 )+2\sqrt{s_1 s_2}}$ and $\Gamma_2=\frac{-i(\omega_1-\omega_2 )}{i(f_1-f_2 )-2\sqrt{s_1 s_2}}$. The square root terms on the right-hand side of Eqn. (2) gives rise to the characteristic distribution of eigenfrequencies with the branching in the parameter space. The Hamiltonian can be solved to obtain two eigenvectors (eigenmodes) $\Phi_\pm$ and a general wavefunction can be defined to describe the temporal evolution of the eigenvectors, $\psi(t)=a_+ \Phi_+$ exp($-i\omega_{+} t)+a_- \Phi_-$ exp($-i\omega_{-} t)$. The amplitudes $a_\pm$ can be solved by considering initial conditions at $t=0$ based on the excitation polarization, for example RCP excitation gives $\psi(0)=\bigl(1/\sqrt{2},  i/\sqrt{2} \bigr)$ and so on. From the wavefunction, one could then calculate the temporal evolution of the spin angular momentum density, field energy density, as well as the DCP as is done for the FDTD simulations. 

Figure~\ref{Fig7} shows the distribution of the absolute difference in the real and imaginary eigenfrequencies calculated using the two-mode approximation model. There are clear signatures of branching and, there are two EPs within the parameter space spanned by Re($\Gamma$) and Im($\Gamma$) as expected from the Hamiltonian. The distribution of the eigenfrequencies do not match with that from FDTD simulations since Re($\Gamma$) and Im($\Gamma$) do not correspond directly to the air hole sweep parameters. Nonetheless, using the model, the eigenmodes in the vicinity of both the EPs are chiral with high DCP but of different handedness. It can be seen from the figure that the handedness of the chirality is independent of the excitation polarization.

\begin{figure*}[t]
\includegraphics[width=0.9 \textwidth]{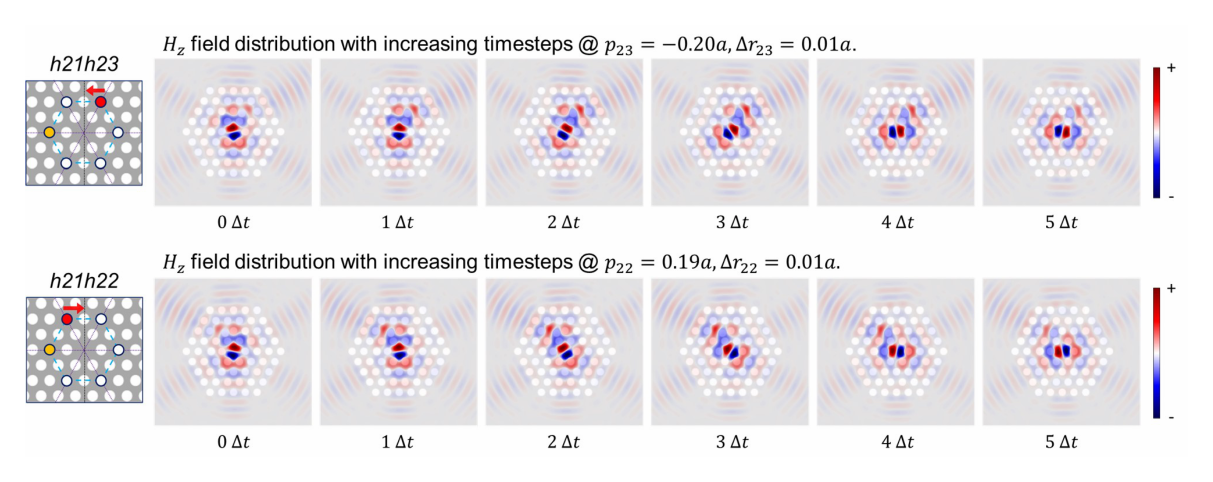}
\caption{
\label{Fig9}Field profiles showing the rotation of the eigenmodes near the EPs for the h21h23 (top) and the h21h22 (bottom) cases. Each timestep $\Delta t$ is $10$ wave period propagation.   
}
\end{figure*}

The temporal evolution of DCP calculated using the two-mode approximation is highly consistent with that from the FDTD simulations (after the excitation has been turned off) for both the RCP [Fig.~\ref{Fig8} (top row)] and LCP [Fig.~\ref{Fig8} (bottom row)] chiral modes. The initial polarization of the cavity fields is essentially determined by the excitation polarization. The cavity fields then proceed to evolve to their intrinsic polarization after the excitation is turned off. The temporal evolution of the polarization is, however, less intuitive for the LP$_{\pm45}$ excitations. After the LP$_{45 (-45)}$ excitation, the cavity fields tend to subsequently become RCP (LCP) regardless of the intrinsic polarization. This suggests that when oriented at $\pm 45^{\circ}$, the linearly polarized excitation imposes a relative $\pm \pi/2$ phase between the two orthogonal linear basis modes before allowing them to reach their intrinsic polarization. As such, for linearly polarized excitation, orientation at $45^{\circ} (-45^{\circ})$ is favored to obtain a larger 〈DCP〉 for the RCP (LCP) chiral modes. Under co-polarized excitation, the instantaneous DCP of the chiral modes remains high throughout the simulation time, giving the maximum $\langle$DCP$\rangle$ which can also be seen in Fig.~\ref{Fig7}. Despite the simplicity of the model, it captures the key features of the FDTD simulation results of the H1 PhC cavity.

\section{Rotating modes}\label{rotatingmodes}

Near the EPs, the eigenmodes rotate in the clockwise (counterclockwise) direction for the h21h23 (h21h22) case as exemplified by the change in the $H_z$ field profile with time in Fig.~\ref{Fig9}.

\section{Further simulation details} \label{sec:furthersimdetails}

We note that our simulations do not consider any material dispersion. Given that we are dealing with resonant modes, the effect of material dispersion is expected to be minimal. Furthermore, for the wavelength of around 1 $\mu$m in GaAs that we are considering here, including the correction for material dispersion~\cite{bliokh2017} should have a negligible effect on the calculation of the DCP.

\begin{acknowledgments}
This work is partly supported by MIC (SCOPE 191503001). Most FDTD simulations for this work are performed on the Supercomputer HOKUSAI BigWaterfall at RIKEN. C.F. Fong is supported by the RIKEN Special Postdoctoral Program. S. Iwamoto acknowledges funding from JST-CREST (JPMJCR19T1) and JSPS KAKENHI (17H06138).
\end{acknowledgments}


%

\end{document}